\documentclass[aps,prl,twocolumn,superscriptaddress]{revtex4}
\usepackage{amsmath}
\usepackage{graphicx}
\usepackage[section]{placeins}
\usepackage{amsmath}
\usepackage{amssymb}
\usepackage{dsfont}
\newcommand{\ket}[1]{|#1\rangle}             % Ket Dirac's notation %
\newcommand{\comment}[1]{}
\begin{document}
\title{Efficient generation and control of different order orbital angular momentum states for communication links}
\author{Sergei Slussarenko}
\affiliation{Dipartimento di Scienze Fisiche, Universit\`{a} di
Napoli ``Federico II'', Compl.\ Univ.\ di Monte S. Angelo, 80126
Napoli, Italy}
\author{Ebrahim Karimi}
\email{karimi@na.infn.it}
\affiliation{Dipartimento di Scienze Fisiche, Universit\`{a} di
Napoli ``Federico II'', Compl.\ Univ.\ di Monte S. Angelo, 80126
Napoli, Italy}
\affiliation{ CNR-INFM Coherentia, Compl.\ Univ.\ di Monte S.
Angelo, 80126 Napoli, Italy}
%%%%
\author{Bruno Piccirillo}
\affiliation{Dipartimento di Scienze Fisiche, Universit\`{a} di
Napoli ``Federico II'', Compl.\ Univ.\ di Monte S. Angelo, 80126
Napoli, Italy}
\affiliation{Consorzio Nazionale Interuniversitario per le Scienze
Fisiche della Materia, Napoli}
%%%%%
\author{Lorenzo Marrucci}
\affiliation{Dipartimento di Scienze Fisiche, Universit\`{a} di
Napoli ``Federico II'', Compl.\ Univ.\ di Monte S. Angelo, 80126
Napoli, Italy}
\affiliation{CNR-SPIN, Compl.\ Univ.\ di Monte S. Angelo, 80126
Napoli, Italy}
%%%%%
\author{Enrico Santamato}
\email{santamato@na.infn.it}
\affiliation{Dipartimento di Scienze Fisiche, Universit\`{a} di
Napoli ``Federico II'', Compl.\ Univ.\ di Monte S. Angelo, 80126
Napoli, Italy}
\affiliation{Consorzio Nazionale Interuniversitario per le Scienze
Fisiche della Materia, Napoli}
\begin{abstract}
We present a novel optical device to encode and decode two bits of information into different Orbital Angular Momentum (OAM) states of a paraxial optical beam. Our device generates the four angular momentum states of order $\pm 2$ and $\pm4$ by Spin-To-Orbital angular momentum Conversion (STOC) in a triangular optical loop arrangement. The switching among the four OAM states is obtained by changing the polarization state of the circulating beam by two quarter wave plates and the two-bit information is transferred to the beam OAM exploiting a single $q$-plate. The polarization of the exit beam is left free for additional one bit of information. The transmission bandwidth of the device may be as large as several megahertz if electro-optical switches are used to change the beam polarization. This may be particularly useful in communication system based on light OAM.
\end{abstract}
\maketitle
%\ocis{(050.1960) Diffraction theory; (260.6042) Singular optics.}
\section{Introduction}
\noindent Dots, dashes and spaces remind the Morse code which were used by sailors, soldiers and postmen in nineteenth century. The electromagnetic waves have been recognized as the fastest way of communication process since Samuel Morse invention, i.e. telegraph. There has been much effort to find a secure and high channel capacity communication process during last two centuries. Most of these links rely on
modulation of intensity (or photon number), frequency, or polarization of the light. Especially, the polarization of an electromagnetic wave associated to the vectorial property of the optical field has been recognized as related to the Spin Angular Momentum (SAM) of a photon and is often proposed as a good candidate for telecommunication and quantum communication protocols~\cite{sergienko}. The photon SAM is inherently binary, so only one bit (in the quantum regime - one qubit) can be encoded over a single photon. Of course, this fact poses limitations to the formulation of quantum communication protocols. Recently, an additional photon degree of freedom associated to the beam phase-front, known as the light orbital angular momentum (OAM), received a great deal of attention for various applications in classical and quantum optics~\cite{allen,molina07}. Each photon of the beam whose state has an azimuthal phase dependance $\exp(im\varphi)$ ($m$ integer) carries a definite amount of OAM equal to $m\hbar$ per photon. In contrast to the SAM, the OAM is inherently multidimensional, and much more information can be encoded into the OAM of a single photon. Such higher-dimension space can be used for expanding the alphabet used in classical and quantum communication~\cite{gibson04} and opens a wide road to many novel protocols not usable for the binary polarization space~\cite{karimipour02}. In the quantum regime, such high order qubits are generally called qudits (with a special case of qutrit for three-level quantum system). Photon qudits have been so far mainly implemented using multi-photon systems or multi-path encoding and the alternative of using OAM encoding has been investigated only in recent times. Up to now, single-photon OAM qudits with dimension $d = 3$ and $d = 4$ have been generated~\cite{torres03, vaziri03} and employed in quantum communication~\cite{molina04}, quantum bit commitment~\cite{Langford04}, and quantum key distribution. However, the difficulty and low efficiency of OAM manipulation has so far represented a serious limitation. In particular, current sources of optical OAM are either very rigid (only one OAM value is generated, with no switching or modulation capability) or very inefficient (typically less than 40\% of the input photons is converted into the desired OAM modes) and fairly expensive; electro-optical fast manipulation of OAM is virtually non-existent, while the OAM control flexibility currently provided by spatial light modulators (SLM) comes at the expense of a slow response ($\sim 1$ kHz) and a high cost.\\

In this paper, we report a fast, reliable, and low-cost device to encode classical (or quantum) information into different OAM states of a light beam. The switching among the OAM states can be realized by electro-optical devices, thus ensuring very fast commutation rate. The beam polarization state is not affected and can be further manipulated to store more information. If also the SAM is considered, our device may encode three classical (or quantum qu-) bits of information into a single photon.
\section{The optical loop device}
The heart of our device is a $q$-plate, a novel optical element made of birefringent liquid crystal spatially oriented in the transverse plane so that it can transfer a well-defined value of topological charge into the output beam depending on the polarization state of the input beam~\cite{marrucci06}. The $q$-plate is characterized by its topological charge $q$ that defines the orientation pattern of the optical axis and its phase retardation $\delta$. When $\delta=\pi$, the $q$-plate is said to be tuned. After tuning, the main effect of the $q$-plate is to convert the SAM of the incident photons into OAM, a process called Spin-To-Orbital angular momentum Conversion (STOC)~\cite{marrucci06}. 
\begin{figure}[h]
	  \begin{center}
	\includegraphics[scale=0.37]{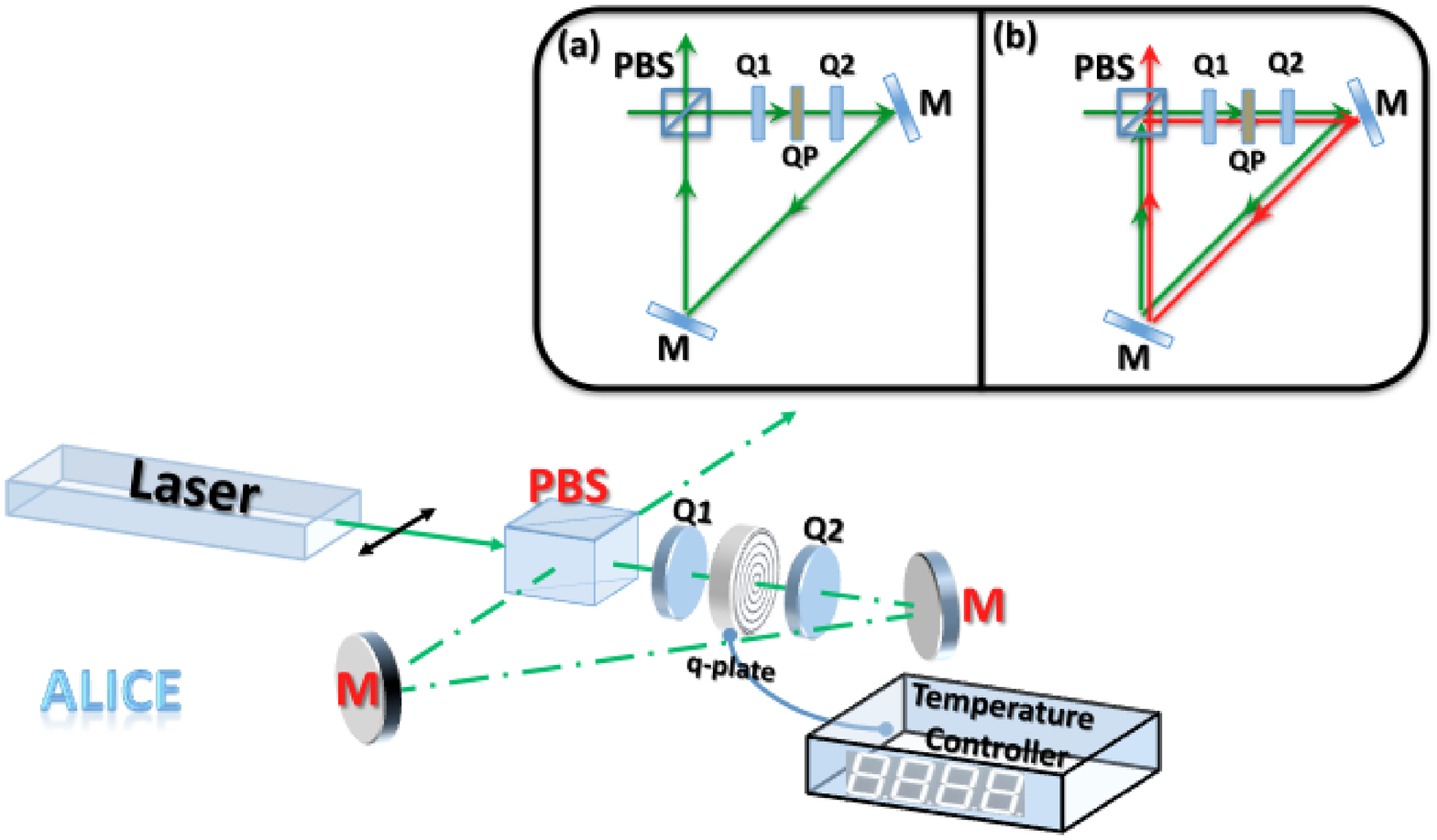}
	\caption{\label{fig:setup} The $q$-plate sandwiched between two QWPs ($Q_1$ and $Q_2$) inserted in the triangular optical loop. The beam trajectory inside the loop device for the case of ($45\ensuremath{^\circ},45\ensuremath{^\circ}$), ($-45\ensuremath{^\circ},-45\ensuremath{^\circ}$) where the beam passes once (inset a), and ($-45\ensuremath{^\circ},45\ensuremath{^\circ}$) and ($45\ensuremath{^\circ},-45\ensuremath{^\circ}$) where the beam passes twice inside the cavity (inset b). The solid green and red lines show the first and second trip, respectively. Legend: M-mirror, PBS-polarizing beam splitter, Q-quarter wave plate, QP-$q$-plate.}
	\end{center}
\end{figure}
The action of the tuned $q$-plate on the incident photon state is described by~\cite{karimi09,karimi09a}
\begin{eqnarray}
	\widehat{\hbox{QP}}\cdot\ket{L,m}=\ket{R,m+2} \nonumber \\
	\widehat{\hbox{QP}}\cdot\ket{R,m}=\ket{L,m-2}
\end{eqnarray}
where $\ket{L}$, $\ket{R}$, and $\ket{m}$ denote the left-circular, the right-circular polarization and the OAM eigenstate with eigenvalue $m$, respectively, and $\widehat{\hbox{QP}}$ is the operator representing the action of the $q$-plate. It is worth nothing that we can change the OAM value of the output photons just by switching the input polarization state, which can be accomplished with frequencies restricted only by electro-optical speed limitations (several MHz).

In our device, the $q$-plate is inserted into a triangular optical loop, as shown in Fig.~\ref{fig:setup}. A polarizing beam-splitter (PBS) is used as the input and output gate, so that only the horizontally polarized light can enter and exit from the loop, the vertically polarized light being directly reflected by the PBS. The tuned $q$-plate was sandwiched between two quarter wave plates (QWPs) and inserted in the loop, as shown in Fig.~\ref{fig:setup}. As it will be shown below, our loop device can
\begin{enumerate}
\item generate the four OAM eigenstates $\ket{\pm2},\ket{\pm4}$; the switch among these four states is made acting on the light polarization so that very fast commutation rate can be achieved;
\item generate qubits formed by any pair sorted from the four OAM eigenstates above; the relative amplitudes of the two states forming the qubit is controlled by acting on the light polarization only;
\item generate a state made of the superposition of all OAM eigenstates with even $m$; the power spectrum of the superposition is controlled by acting on the light polarization only.
\end{enumerate}
\section{Pure OAM eigenstates generation}
Let us consider a TEM$_{00}$ laser beam with OAM $m=0$ entering in the optical loop. The output beam is a pure state of order $|m|=2, 4$ when the QWPs in the loop are set at $\pm45\ensuremath{^\circ}$. Let us consider, for example, the case where the two QWPs were set at $45\ensuremath{^\circ}$. The first QWP changes the polarization of the beam circulating in the loop from the horizontal ($\ket{H}$) to the left circular ($\ket{L}$). The $q$-plate coherently transfers Spin-to-OAM, switches the polarization into the right circular ($\ket{R}$) and provides the beam with an OAM value $m=+2$. The second QWP switches back the right-circular polarization into the horizontal one ($\ket{H}$), so that the light was led out from the loop. Because of the even number of reflections by mirrors, the OAM of the output beam is left to $m=+2$. The full sequence of changes of the photon state is
\begin{eqnarray}
   \ket{H,0}\stackrel{\scriptsize{Q}_1^{@45\ensuremath{^\circ}}}{\longrightarrow}
   \ket{L,0}\stackrel{\scriptsize{QP}}{\longrightarrow}
   \ket{R,2}\stackrel{\scriptsize{Q}_2^{@45\ensuremath{^\circ}}}{\longrightarrow}\ket{H,2}
\end{eqnarray}
The same process occurs with the two QWPs set at $-45\ensuremath{^\circ}$. In this case, however, the output beam is left with $m=-2$. The full sequence is
\begin{eqnarray}
   \ket{H,0}\stackrel{\scriptsize{Q}_1^{@-45\ensuremath{^\circ}}}{\longrightarrow}
   \ket{R,0}\stackrel{\scriptsize{QP}}{\longrightarrow}
   \ket{L,-2}\stackrel{\scriptsize{Q}_2^{@-45\ensuremath{^\circ}}}{\longrightarrow}\ket{H,-2}
\end{eqnarray}
Figure \ref{fig:setup}-(a) shows the ray trajectory inside the optical loop for these two cases.\\

\begin{figure}[!htbp]
	\begin{center}
	\includegraphics[scale=0.5]{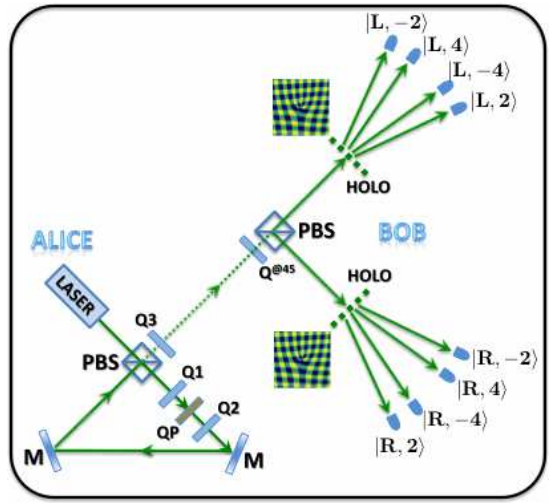}
	\caption{\label{fig:8D_setup} Alice apparatus is similar to what has already shown in figure~(\ref{fig:setup}). In order to encode another information bit, a third QWP ($Q_3$) is located at the exit face of the optical loop. Alice is able to generate $(\ket{L,-4},\ket{L,-2},\ket{L,+2},\ket{L,+4},\ket{R,-4},\ket{R,-2},\ket{R,+2}$ and $\ket{R,+4})$ by setting her QWPs at $\pm45\ensuremath{^\circ}$. Bob measures the photon SAM state by suitable QWP and a PBS and measures the photon OAM state by suitable holograms, as shown.}
	\end{center}
\end{figure}

When the two QWPs are set at opposite angles ($45\ensuremath{^\circ},-45\ensuremath{^\circ}$) or ($-45\ensuremath{^\circ},45\ensuremath{^\circ}$), the output beam is left with $m =\pm4$, respectively. Let us consider, for example, the case where the first QWP is set at $+45\ensuremath{^\circ}$ and the second at $-45\ensuremath{^\circ}$, respectively. The horizontal polarized beam circulating in the optical loop is changed into the left-circular polarization by the first QWP. The $q$-plate, then, coherently transfers the spin into OAM and the state changes into $\ket{R,2}$. The second QWP switches back the polarization into the vertical polarization state, so that the beam is reflected back into the loop by the PBS. However, the sign of the OAM changes due to the odd number of reflections by mirrors and PBS. In the second trip, the first QWP changes the vertical polarization into the right-circular polarization and the $q$-plate transfers the polarization state into the left-circular polarization and subtracts $2$ to the the beam OAM leading to $m=-4$. After that, the second QWP changes back the left-circular polarization into the horizontal polarization so that the beam with $m=-4$ can leave the loop after an even number of reflections by mirrors and PBS. For the ($-45\ensuremath{^\circ},45\ensuremath{^\circ}$) configuration the same process takes place, but the sign of the output OAM is reversed. Inset (b) in Fig.~\ref{fig:setup}-(b) shows the ray trajectory inside the optical loop for the last two cases. The full sequences of changes of the photon states are ($M$ represents here the two mirrors)
\begin{eqnarray} \label{eq:l4}
  	   \ket{H,0}&\stackrel{\scriptsize{Q}_1^{@45\ensuremath{^\circ}}}{\longrightarrow}&
       \ket{L,0}\stackrel{\scriptsize{QP}}{\longrightarrow}
       \ket{R,2}\stackrel{\scriptsize{Q}_2^{@-45\ensuremath{^\circ}}}{\longrightarrow}
       \ket{V,2}\stackrel{\scriptsize{M+PBS}}{\longrightarrow}
       \cr
       &&
       \ket{V,-2}\stackrel{\scriptsize{Q}_1^{@45\ensuremath{^\circ}}}{\longrightarrow}
       \ket{R,-2}\stackrel{\scriptsize{QP}}{\longrightarrow}
       \ket{L,-4}\stackrel{\scriptsize{Q}_2^{@-45\ensuremath{^\circ}}}{\longrightarrow}\ket{H,-4}
       \cr\cr
       \ket{H,0}&\stackrel{\scriptsize{Q}_1^{@-45\ensuremath{^\circ}}}{\longrightarrow}&
       \ket{R,0}\stackrel{\scriptsize{QP}}{\longrightarrow}
       \ket{L,-2}\stackrel{\scriptsize{Q}_2^{@-45\ensuremath{^\circ}}}{\longrightarrow}
       \ket{V,-2}\stackrel{\scriptsize{M+PBS}}{\longrightarrow}
       \cr
       &&
       \ket{V,2}\stackrel{\scriptsize{Q}_1^{@45\ensuremath{^\circ}}}{\longrightarrow}
       \ket{L,2}\stackrel{\scriptsize{QP}}{\longrightarrow}
       \ket{R,4}\stackrel{\scriptsize{Q}_2^{@-45\ensuremath{^\circ}}}{\longrightarrow}\ket{H,4}
\end{eqnarray}
Therefore, the loop device is able to generate $-4$,$-2$,$+2$,$+4$ values of OAM by choosing the proper angles for the two QWPs. Table (\ref{table}) shows the four possible combinations of QWP angles and the corresponding OAM values of the output beam~\cite{note01}.
\begin{table}[!htbp]
  \centering
  \caption{\label{table} Four possible combinations of
QWP angles and their corresponding beam's OAM values.}
  \begin{tabular*}{0.48\textwidth}{@{\extracolsep{\fill}} c|ccc}
  \hline\hline
  Logical bit & Q$_{1}$  &  Q$_{2}$  & OAM value  \\ \hline
  $00$    & $+45\ensuremath{^\circ}$  &  $+45\ensuremath{^\circ}$  & $+2$   \\
  $01$    & $-45\ensuremath{^\circ}$  &  $-45\ensuremath{^\circ}$  & $-2$   \\
  $10$    & $+45\ensuremath{^\circ}$  &  $-45\ensuremath{^\circ}$  & $-4$   \\
  $11$    & $-45\ensuremath{^\circ}$  &  $+45\ensuremath{^\circ}$  & $+4$   \\
  \hline\hline
  \end{tabular*}
\end{table}
One may replace the QWP with electro-optical devices so to encode the information in the light beam with rate of the order of several megahertz.\\
The optical loop setup proposed in this work can be used for classical communications in 8D SAM-OAM space. As we have already mentioned, an additional classical bit can be encoded in the SAM of the output beam by inserting a further QWP at the exit of the optical loop. So, Alice can transmit to Bob the eight spinorbit photon states $(\ket{L},\ket{R})\otimes(\ket{-4},\ket{-2},\ket{+2},\ket{+4})$, corresponding to three bit of information per photon. Bob can use, for example, a QWP at $45\ensuremath{^\circ}$ followed by a PBS to select the SAM state of the received photons and the holograms shown in Fig.~(\ref{fig:8D_setup}) to discriminate the photon OAM~\cite{gibson04}. The communication transmitter and receiver scheme shown in Fig.~(\ref{fig:8D_setup}) can be fully realized by the available technology. Its main advantage is that three bits are encoded in each photon manipulating only the polarization degree of freedom, which can be achieved by very fast and efficient electro-optical switching. Our apparatus has been intended for classical telecommunication, but it can be applied for single photon quantum communication too, since the $q$-plate can act as a quantum device~\cite{nagali09a,nagali09b}.
\section{OAM qubit generation}
When the orientation angles of the QWPs are set to values different from those reported in Table~(\ref{table}), a superposition of OAM states is generated, in general. In this case, the light is trapped inside the cavity making infinite number of loops. The output state is then given by a superposition of different OAM eigenstates made by the portions of horizontally polarized light exiting the cavity at each loop. In the quantum regime, the superposition is among the probability amplitudes $\alpha_N$ that the photon exits the optical loop after $N$ round trips. When the angle of one of the QWPs inside the cavity is fixed at $45\ensuremath{^\circ}$ (or $-45\ensuremath{^\circ}$), four different qubits are produced made of any two of the four OAM states $\ket{\pm2},\ket{\pm4}$. \\ More precisely, if the first (second) QWP is fixed at angle $45\ensuremath{^\circ}$ the generated output state up to a global phase factor is given by
\begin{equation}\label{eq:qubit1}
   \ket{\psi_1}=C_1(\theta,\psi)[2(\cos(2\theta+\psi)-\sin\psi)\ket{2}-
                       i(1-\sin2\theta)\ket{\mp4}]
\end{equation}
where $C_1(\theta,\psi)$ is a normalization factor depending on the round trip phase delay $\delta$ and on the orientation angle $\theta$ of the free QWP.\\
If the first (second) QWP is fixed at $-45\ensuremath{^\circ}$, instead, the output state is given by
\begin{equation}\label{eq:qubit2}
   \ket{\psi_2}=C_2(\theta,\psi)[2(\cos(2\theta-\psi)-\sin\psi)\ket{-2}-
                       i(1+\sin2\theta)\ket{\pm4}]
\end{equation}
where $C_2(\theta,\psi)$ is a new normalization factor. It is worth noting that the relative phase of the two OAM eigenstates forming the qubit is fixed to be $\pm 90\ensuremath{^\circ}$, so that only the relative amplitude can be changed by the control parameters $\theta$ and $\psi$. The possibility of exploiting the photon polarization to control qubits formed by two OAM eigenstates with different $m$ may be useful for quantum computing or other quantum applications.
\section{Multiple OAM generation}
\begin{figure}[!htbp]
	\begin{center}
	\includegraphics[scale=0.3]{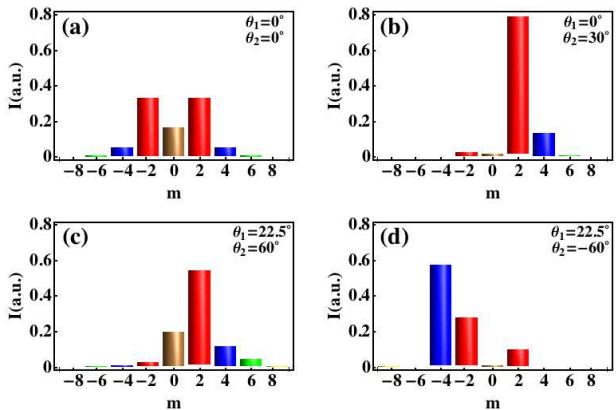}
	\caption{\label{fig:oam_super} Calculated OAM power spectrum $I_{n}=|c_{n}|^{2}$ of the beam emerging form the loop device for different angles $\theta_1$ and $\theta_2$ of the two QWPs for loop delay $\psi=0$. The power spectrum can be either symmetric (a) or not symmetric (b,c,d) and the fundamental $m=0$ component can be suppressed (b,d).}
	\end{center}
\end{figure}

When both the angles of the QWPs are different from $\pm45\ensuremath{^\circ}$, a complex superposition of even OAM eigenstates is generated, having the general form $\sum_{n=-\infty}^{+\infty} c_{|2n|}\ket{2n}$, where $c_{|2n|}$ depend on the angles $\theta_1$ and $\theta_2$ of the two QWPs and on the loop delay $\psi$. Figure~\ref{fig:oam_super} shows some examples of infinite OAM state superposition obtained for different orientations $\theta_1$ and $\theta_2$ of the two QWPs and for $\psi=0$. Notice how the symmetry of the OAM power spectrum of the output beam is strongly affected by $\theta_1$ and $\theta_2$. The odd OAM components are missing because we used a $q=1$ $q$-plate. A full OAM specrum can be generated by using a $q=1/2$ q-plate.  The possibility of exploiting the light polarization to control full spectra of OAM eigenstates may be useful for future, yet not identified, applications.
\section{The experiment}
\begin{figure}[!htbp]
	\begin{center}
	\includegraphics[scale=0.5]{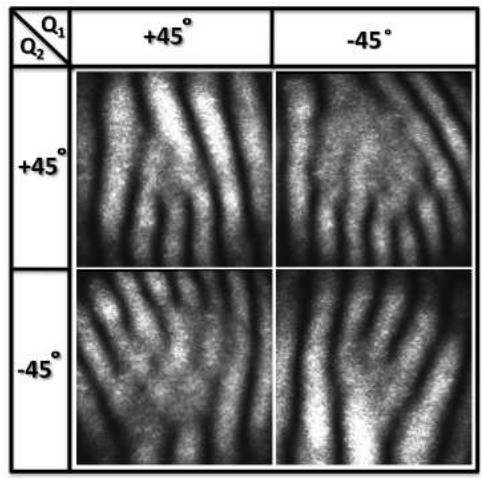}
	\caption{\label{fig:interference} The interference of the output beam from the loop device and TEM$_{00}$ beam for the four QWP angles shown in Table~(\ref{table}).}
	\end{center}
\end{figure}

\begin{figure}[!htbp]
	\begin{center}
	\includegraphics[scale=0.6]{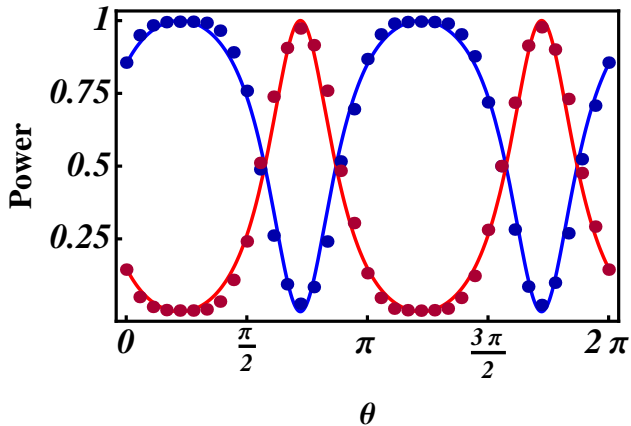}
	\caption{\label{fig:qubit} The normalized powers of the $m=2$ (red) and $m=-4$ (blue) components of the loop output beam as functions of the orientation angle $\theta$ of the second QWP. The first QWP was held fixed at 45\ensuremath{^\circ}. The continuous curve is the fit to Eq.~(\ref{eq:qubit1}. The optical retardation $\psi$ of the loop was used as fitting parameter. In the case of the figure above we found a best fit value $\psi=0$. }
	\end{center}
\end{figure}

In our first experiment, we used a c.w. TEM$_{00}$ laser source at $\lambda=532$~nm and measured the output beam phase-front by making an interference with a plane-like phase-front of same frequency. We used an azimuthally oriented liquid crystal home made $q$-plate. The optical retardation of the $q$-plate was tuned by temperature controller~\cite{karimi09} in such a way that it acted as a half-wave plate ($\delta=\pi$). Figure~(\ref{fig:interference}) shows the recorded interference pattern of the beam exiting the optical loop for different angles of the QWPs. The absolute value of the OAM is deduced from the number of prongs of the interference fork and the sign from the prongs up or down direction.
In our second experiment, we fixed the first QWP at $45\ensuremath{^\circ}$ and rotated the second one so to generate the qubit formed by the OAM eigenstates $2$ and $-4$ as described before. The alignment of the loop were adjusted by moving the two mirrors so to obtain a good symmetric interference pattern~\cite{note02}. For each angle $\theta$ of the second QWP, we measured the power flow associated to the $m=2$ and $m=-4$ components of the beam exiting the loop device by suitable computer generated fork holograms displayed onto a Spatial Light Modulator (SLM). Beyond the hologram, the $m=0$ component was selected by a pinhole posed in the focal plane of a convergent lens. Finally, the measured power flows were normalized so that their summation returned one. The result is shown in Fig.~\ref{fig:qubit}. The full curve is from Eq.~(\ref{eq:qubit2}). To fit the data we used the loop retardation $\psi$ as best fit parameter.
%
%%%%%%%%%%%% Add the new experiment with curve
\section{Conclusions}
We presented a loop device based on $q$-plate to generate and encode two bits information into the OAM of a single photon. The encoding process is very efficient (nominal efficiency is 100\%) and very fast, because it can be fully implemented by electro-optical devices. The encoded information can be read with a computer generated hologram properly designed to detect all four OAM states simultaneously~\cite{gibson04}. The generation process is deterministic and the setup is suitable for both classical and quantum regimes of light. Furthermore, the optical loop can be easily modified to encode three bits of information in a single photon by adding an additional polarization bit. The same setup allows also the generation of qubits made of two different OAM orders or qudits with infinite number of OAM eigenstates. The generation process of single OAM eigenstates, OAM qubits and OAM qudits with $d=\infty$ is deterministic, has nominal 100\% efficiency, and the output OAM state can be controlled by very fast electro-optical devices.


\begin{thebibliography}{99}
\bibitem{note01} 
Because we are only interested to the OAM value of the beam, we have neglected the phase
delay between the first and second trip inside the optical loop. Indeed, the beam with OAM
value |m| = 4 has a trip delay with respect to the beam with OAM value |m| = 2.
\bibitem{note02} 
This pattern has the characteristic form of six-petal flower.
\end{thebibliography}
\end{document}